\documentclass{JHEP3}\newcommand{\format} {\JHEPformat}
\usepackage{amsmath}
\usepackage{amssymb}
\usepackage{epsfig}
\usepackage{latexsym}

\newcommand{\JHEPformat} {
\bibliographystyle{JHEP}
\newcommand{\maketitlepage} {}
\abstract{\theabstract}
\keywords{\thekeywords}
\preprint{\thepreprint}
}

\newcommand{\TITLE}[1] {\newcommand{\thetitle} {#1}\title{#1}}
\newcommand{\ABSTRACT}[1] {\newcommand{\theabstract} {#1}}
\newcommand{\AUTHOR}[1] {\newcommand{\theauthor} {#1}}
\newcommand{\ADDRESS}[1] {\newcommand{\theaddress} {#1}}
\newcommand{\DATE}[1] {\newcommand{\thedate} {#1}\date{#1}}
\newcommand{\KEYWORDS}[1] {\newcommand{\thekeywords} {#1}}
\newcommand{\PREPRINT}[1] {\newcommand{\thepreprint} {#1}}

\newcommand{\N}{{\cal N}}
\newcommand{\HH}{{\cal H}}
\newcommand{\cs}{{\cal S}}
\newcommand{\e}{{\epsilon}}
\newcommand{\s}{{\sigma}}
\newcommand{\ka}{{\kappa}}
\newcommand{\K}{{\rm K}}
\newcommand{\One} {{\bf 1}} 
\newcommand{\tr} {\operatorname{tr}}
\newcommand{\sm} {{\scriptstyle{-}}}
\newcommand{\mat}[1] {\begin{pmatrix}#1\end{pmatrix}}
\newcommand{\bra}[1] {\left<#1\right|}
\newcommand{\ket}[1] {\left|#1\right>}
\newcommand{\braket}[2] {\left<#1\vphantom{#1}\right|
                         \left.\!\vphantom{#1}{#2}\right>}
\newcommand{\brao}[1] {\left(#1\right|}
\newcommand{\keto}[1] {\left|#1\right)}
\newcommand{\braketo}[2] {\left(#1\vphantom{#1}\right|
                         \left.\!\vphantom{#1}{#2}\right)}

\TITLE{Squeezed State Projectors in String Field Theory}

\AUTHOR{{\bf Ehud Fuchs}\footnote{\tt udif@tau.ac.il},
{\bf Michael Kroyter}\footnote{\tt mikroyt@tau.ac.il},
{\bf Alon Marcus}\footnote{\tt alon@tau.ac.il}}

\ADDRESS{School of Physics and Astronomy\\
  The Raymond and Beverly Sackler Faculty of Exact Sciences\\
  Tel Aviv University, Ramat Aviv, 69978, Israel
}

\author{Ehud Fuchs, Michael Kroyter, Alon Marcus\\
School of Physics and Astronomy\\
The Raymond and Beverly Sackler Faculty of Exact Sciences\\
Tel Aviv University, Ramat Aviv, 69978, Israel\\
E-mails:
\email{udif@tau.ac.il}, \email{mikroyt@tau.ac.il}, \email{alon@tau.ac.il}
}

\ABSTRACT{
We find a new subalgebra of the star product in the matter sector.
Its elements are squeezed states whose matrices commute with $(\K_1)^2$.
This subalgebra contains a large set of projectors.
The states are represented by their eigenvalues and we find a mapping
between the eigenvalues representation and other known representations.
The sliver is naturally in this subalgebra.
Surprisingly, all the generalized butterfly states are also
in this subalgebra, enabling us to analyze their spectrum,
and to show the orthogonality of different butterfly states.
This means that multi $D$-brane states can be built of
butterfly states.
}

\DATE{June 2002}
\PREPRINT{TAUP-2704-02\\{\tt hep-th/0207001}}

\format

\begin{document}

\maketitlepage

\section{Introduction and summary}

Cubic string field theory \cite{Witten:1986cc}  has attracted renewed interest,
mainly due to Sen's conjecture on tachyon condensation \cite{Sen:1999xm}.
For bosonic string theory the conjecture is that the open string tachyon
potential has a minimum that the tachyon condenses to. The value
of the potential at this minimum is exactly equal to the tension of
a $D$-brane.
The tachyon potential was calculated numerically in cubic open string
field theory using the level truncation scheme
\cite{Kostelecky:1990nt,Sen:1999nx,Taylor:2000ek,Moeller:2000xv,deMelloKoch:2000ie}.
These calculations agree very well with Sen's conjecture.

Kostelecky and Potting attempted in {\cite{Kostelecky:2000hz} to find
the vacuum of string field theory analytically.
The basis for their solution in the matter sector was a solution to the
projection equation
\begin{equation}
\label{proj}
\cs=\cs\star\cs\,,
\end{equation}
where the product is the star product of cubic string field theory
\cite{Witten:1986cc}. They assumed that the solution is a
squeezed state of the form
\begin{equation}
\label{squeezed}
\ket{\cs}=\exp\left(-\frac{1}{2}\sum_{n,m=1}^{\infty}
    a_n^\dagger S_{nm} a_m^\dagger\right)\ket{0}\,.
\end{equation}
They made one more simplifying ansatz
\begin{equation}
\label{KPansatz}
[T,V^{rs}]=0\,,\qquad{\scriptstyle r,s=1\ldots3}\,,
\end{equation}
where $T=CS$, the twist matrix $C_{nm}=(-1)^m\delta_{nm}$
and $V^{rs}$ are the three-vertex matrices \cite{Gross:1987ia}.
These assumptions allow for only two solutions.
One trivial solution which is the identity state $S=C$
and one non-trivial solution.
A little earlier the subalgebra of wedge states was found
\cite{Rastelli:2000iu}.
Two of those states are projectors:
the $360^\circ$ wedge, which corresponds to the identity state
and the infinitely thin wedge called the sliver \cite{Rastelli:2001jb},
which corresponds to the non-trivial solution of \cite{Kostelecky:2000hz}.

It seems surprising that the first two non-trivial projection states
found, turned out to be the same state,
especially due to the fact that in \cite{Rastelli:2000iu}
CFT techniques were used while \cite{Kostelecky:2000hz} used oscillators.
The reason oscillator based calculations gave a wedge
state is that the three-vertex matrices are related to the $120^\circ$
wedge state and all wedge state matrices commute. Therefore the ansatz
(\ref{KPansatz}) restricts the solutions to wedge state solutions.

Projectors are even more relevant in vacuum string field theory (VSFT),
which is a formulation of SFT around the tachyon vacuum
\cite{Rastelli:2000hv,Rastelli:2001vb,Rastelli:2001uv,Gaiotto:2001ji}.
In this formulation the kinetic operator ${\cal Q}$ is purely ghost,
meaning that the equation of motion for the matter sector is simply
the projection equation (\ref{proj}).
Therefore, projectors are solitonic solutions of VSFT associated
with $D$-branes.
The expectation that there will be only one type of $D25$-brane does not
agree with the infinite number of spatially independent rank-one projectors.
This suggests that all these projectors are related by a gauge transformation.

Rastelli Sen and Zwiebach found the spectrum of the three-vertex
and wedge states matrices \cite{Rastelli:2001hh}.
First the eigenvalues and eigenvectors of the matrix $\K_1$
were calculated, where the matrix $\K_1$ is defined as the action of the
star algebra derivation $K_1=L_1+L_{-1}$ in the oscillator basis.
The eigenvalues of $\K_1$ are continuous in the range $-\infty<\ka<\infty$
and are non-degenerate.
$\K_1$ also satisfies the commutation relations
\begin{equation}
\label{K1Commute}
[M^{rs},\K_1]=[T_N,\K_1]=0\,,
\qquad{\scriptstyle r,s=1\ldots3}\,,\quad{\scriptstyle N=1\ldots\infty}\,,
\end{equation}
where $V^{rs}=CM^{rs}$ are the three-vertex matrices and
$V_N=CT_N$ are the wedge state matrices.
The eigenvectors of $\K_1$ are the eigenvectors of $M^{rs},T_N$
because $\K_1$ is non-degenerate and eq.~(\ref{K1Commute}).
The spectroscopy results simplified many elaborate computations.
It was used in \cite{Douglas:2002jm} to formulate
the continuous Moyal representation of the star algebra, in
\cite{Imamura:2002rn} to study the gauge transformation of the vector state 
in VSFT, in \cite{Okuyama:2002tw} for an analytical calculation of tensions
ratio, and in \cite{Okuyama:2002yr} for proving the equivalence of two
definitions of ${\cal Q}$.

The spectroscopy simplifies the calculations of
\cite{Kostelecky:2000hz} as well.
We can work in the $\K_1$ basis where the $M^{rs}$ matrices are diagonal,
and require that $T=CS$ should also commute with $\K_1$
\begin{equation}
\label{KPanalog}
[T,\K_1]=0\,.
\end{equation}
Instead of equations involving infinite matrices, we get
scalar equations for each eigenvalue $\ka$.
The condition (\ref{KPanalog}) is exactly analog to the ansatz
(\ref{KPansatz}). Thus, repeating the calculations of
\cite{Kostelecky:2000hz}, using the spectroscopy results, gives
the same solutions.

The placement of the factors of $C$ in the above commutation relations is
very rigid due to the fact that $C$ does not commute with $\K_1$.
The main idea of this paper is to rely on the commutation relation
\begin{equation}
[C,(\K_1)^2]=0\,.
\end{equation}
This is a result of the double degeneracy of $(\K_1)^2$ where each
eigenvalue $\ka^2$ has two eigenvectors $v^{(\pm\ka)}$.
We solve the equations of \cite{Kostelecky:2000hz} using a weaker ansatz
\begin{equation}
\label{TheAnsatz}
[S,(\K_1)^2]=0\,.
\end{equation}
States that satisfy this ansatz form a subalgebra $\HH_{\ka^2}$ of
the star product.
States in this subalgebra are block diagonal in the $\K_1$ basis
and are represented using two by two matrices.
The advantage of the weaker ansatz is that now we get a larger
set of solutions. The calculations and the solutions are presented in
section \ref{projectors}.

Understanding the meaning of our solutions requires translating them into
more familiar representations. This is done in section \ref{representations}.
It turns out that the entire family of generalized butterfly states
\cite{Gaiotto:2001ji,Schnabl:2002ff,Gaiotto:2002kf}
is in the $\HH_{\ka^2}$ subalgebra and we find their spectra (\ref{gens}).

The fact that the butterfly state
\begin{equation}
\exp\left(-\frac{1}{2}L_{-2}\right)\ket{0}=\exp\left(-\frac{1}{2}a^\dagger V^B a^\dagger\right)\ket{0}\,,
\end{equation}
satisfies our ansatz, meaning $[V_B,(\K_1)^2]=0$, was somewhat unexpected.
An explicit calculation of the butterfly spectroscopy can be found
in the appendix.
It seems that other projectors with a simple Virasoro structure,
do not satisfy our ansatz.

Not all our solutions correspond to surface states.
One interesting such solution is the dual of the nothing state
\begin{equation}
\ket{\cs}=\exp\left(-\frac{1}{2}a^\dagger (-I) a^\dagger\right)\,.
\end{equation}
It looks like the nothing state only with an opposite sign in the exponent.
The nothing state describes a configuration of a string 
with an $X_n$ independent wave function, that is
$\Phi[\pi_k]=\prod_{n=1}^\infty\delta(\pi_n)$, where $\pi_n$ are the conjugate
momenta.
The dual of the nothing is
$\Phi[X_k]=\prod_{n=1}^\infty\delta(X_n)$.

In section \ref{Moyal} we discuss the $\HH_{\ka^2}$ subalgebra
in the Moyal representation.
We show that all our projectors have the same normalization
in this formalism and show the symmetry that relates them.
We also show that the generalized butterfly states are orthogonal.
This result can be used to build multi $D$-brane states from squeezed
states only, without the need of non Gaussian states.

In the rest of the paper states are written up to normalization.
The singular normalization of string field states
is supposed to be corrected by the ghost sector, which 
we did not treat in this paper.
We also ignored the zero modes, assuming that the string field is
independent of them, meaning that our solutions are analog to $D25$-branes.
The spacetime index $\mu=0\ldots25$, is also suppressed.
One can use the spectroscopy of the matter sector including the zero modes,
and that of the ghost sector
\cite{Feng:2002rm,Belov:2002fp,Potting:2002jh,Erler:2002nr},
to generalize this work.

\section{Projectors in the $\HH_{\ka^2}$ subalgebra}
\label{projectors}

Squeezed states whose matrices commute with $(\K_1)^2$ form a subalgebra
of the star product, which we denote $\HH_{\ka^2}$.
To prove this statement we write the expression of the star product of two
squeezed states,
$\ket{\cs_3}=\ket{\cs_1}\star\ket{\cs_2}$ using \cite{Kostelecky:2000hz}
\begin{equation}
\label{KP3eq}
CS_3C=V^{11}+\mat{V^{12},& V^{21}}\cdot
         \mat{\One-S_1 V^{11} & \phantom{\One}-S_1 V^{12}\\
              \phantom{\One}-S_2 V^{21} & \One-S_2 V^{11}}^{-1}
    \cdot\mat{S_1 V^{21} \\ S_2 V^{12}}\,,
\end{equation}
where $V^{rs}$ are the three-vertex matrices.
In~\cite{Rastelli:2001hh} it was shown that
\begin{equation}
\label{McomK}
[M^{rs},\K_1]=0\,,
\end{equation}
where $M^{rs}=CV^{rs}$.
The fact that $[C,(\K_1)^2]=0$ completes the proof,
since $S_3$ is a function of matrices that
commute with $(\K_1)^2$, and therefore $[S_3,(\K_1)^2]=0$.

Eq.~(\ref{McomK}) with the nondegeneracy of $\K_1$ implies
that the $M^{rs}$ matrices are
diagonal in the $\K_1$ basis.
In~\cite{Rastelli:2001hh} their eigenvalues were found
\begin{equation}
\label{mu}
\begin{split}
\mu(\ka)\equiv\mu^{11}(\kappa)&=-\frac{1}{1+2\cosh(\kappa \pi/2)}\,,\\
\mu^{12}(\kappa)&=\frac{1+\exp(\kappa \pi/2)}
   {1+2\cosh(\kappa \pi/2)} \,,\\
\mu^{21}(\kappa)&=\mu^{12}(-\kappa)\,.
\end{split}
\end{equation}

To find squeezed state projectors one has to solve eq.~(\ref{KP3eq}),
setting $S_1=S_2=S_3$.
In the $\HH_{\ka^2}$ subalgebra the projector condition
becomes a set of equations consisting of one scalar equation for $\ka=0$
and one $2\times 2$ matrix equation for each $\ka >0$.
We shall now solve this set of equations for all $\ka$ to find
the condition for a state in the $\HH_{\ka^2}$ subalgebra to be a projector.

\subsection{The $\kappa=0$ subspace}
\label{ka=0}
For the $\ka=0$ eigenvalue there is a single normalizable eigenvector
of $(\K_1)^2$.
We use the fact that it is twist odd to set
$V^{rs}=-\mu^{rs}(0)$ in (\ref{KP3eq}) and get
\begin{equation}
\label{KPk0}
s_0=\frac 1 3+\mat{-\frac 2 3, & -\frac 2 3}
  \mat{1-\frac 1 3 s_0 & \phantom{1-{}}\frac 2 3 s_0 \\
       \phantom{1-{}}\frac 2 3 s_0 & 1-\frac 1 3 s_0 }^{-1}
 \mat{s_0 & 0 \\
      0 & s_0}
 \mat{-\frac 2 3 \\ -\frac 2 3}\,.
\end{equation}
The matrix that has to be inverted  has an inverse for $s_0 \neq 1,-3$.
The condition $s_0 \neq 1$ is related to the fact that $S$ defines
a Bogoliubov transformation, and thus the eigenvalues of $S^\dagger S$
should obey
\begin{equation}
\label{BogConst}
\lambda_{S^\dagger S}<1\,.
\end{equation}
We know, however, that projectors of SFT are singular. When
the matrix is inverted eq. (\ref{KPk0}) gives
\begin{equation}
\label{s0Solutions}
s_0=\pm 1\,,
\end{equation}
We see that although singular,
$s_0=1$ is a solution to the equation.
Indeed this is the solution for all surface state projectors
with $f(\pm i)=\infty$, where $f(z)$ is the canonical transformation defining
the state, as was shown in \cite{Gaiotto:2002kf}.
The $s_0=-1$ solution can represent either a non-surface-state
projector, or a surface state for which $f(\pm i)\neq\infty$,
as in the case of the star-algebra identity $S=C$, for which
\begin{equation}
\label{Identity}
f(z)=\frac{z}{1-z^2}.
\end{equation}

\subsection{The $\kappa \neq 0$ subspace}

The eigenvalue $\ka^2$ of $(\K_1)^2$ is doubly degenerate 
for $\ka\neq0$ with the eigenvectors $v^{(\pm\ka)}$.
To solve the projection equation for the $\pm\ka$ pairs 
we need to work in the two dimensional subspace spanned by
\begin{equation}
\mat{1\\0}\equiv v^{(-\ka)}\qquad\mat{0\\1}\equiv v^{(\ka)}\,.
\end{equation}
We avoid a double counting by taking only $\kappa>0$.
In this subspace the entities of eq. (\ref{KP3eq}) are two by two matrices.
The action of the $C$ matrix in this subspace is
\begin{equation}
C_{\ka}=\mat{\phantom{-}0 & -1\\-1 & \phantom{-}0}\,,
\end{equation}
since the vectors
$v_\pm^{(\ka)}=\frac{1}{2}\left(v^{(-\ka)}\mp v^{(\ka)}\right)$~\cite{Rastelli:2001hh}
are eigenvectors of $C$ with eigenvalues $\pm 1$.
Using eq. (\ref{mu}) we can now write the 3-vertex in this subspace
\begin{equation}
\label{V3}
\begin{split}
V^{11}&=\frac{1}{1+2\cosh(\kappa \pi/2)}
\mat{0 & 1 \\
     1 & 0}\,,\\
V^{12}&=\frac{-1}{1+2\cosh(\kappa \pi/2)}
\mat{0 & 1+\exp(\kappa \pi/2) \\
     1+\exp(-\kappa \pi/2) & 0}\,,\\
V^{21}&=(V^{12})^T\,.
\end{split}
\end{equation}
In this subspace a state in the $\HH_{\ka^2}$ subalgebra is given by
\begin{equation}
\label{Sansatz}
S_{\ka}=
\mat{ s_1(\ka) & s_2(\ka) \\
      s_2(\ka) & s_3(\ka) } \,.
\end{equation}
$S_{\ka}$ is symmetric because it represents a quadratic form.

The functions $s_1,s_2,s_3$ are complex in general, but we should impose
the BPZ reality condition on the string field state.
The condition is that the BPZ conjugate of the state
should be equal to the hermitian conjugate of the state
\cite{Witten:1986cc,Witten:1986qs,Belavin:1984vu}
\begin{equation}
\label{RealState}
\braket{V_2}{\Psi} \equiv \bra{\Psi}_{\text{BPZ}}
=\left(\ket{\Psi}\right)^\dagger\,.
\end{equation}
For squeezed states the reality condition translates into the condition
\begin{equation}
CSC=S^\dagger\,,
\end{equation}
and for the $\HH_{\ka^2}$ subalgebra we get the conditions
\begin{equation}
\label{SReal}
\begin{gathered}
s_0=s_0^*\,,\\
\mat{s_3 & s_2\\s_2 & s_1}=\mat{s_1^* & s_2^*\\s_2^* & s_3^*}
\Rightarrow \begin{cases}s_3=s_1^*\\s_2=s_2^*\end{cases}\,.
\end{gathered}
\end{equation}

Solving the projection equation we get two valid
solutions (before imposing any reality condition)
\begin{gather}
\label{OneSolution}
s_1=s_3=0,\ \ \ s_2=-1\,.\\
\label{ManySolutions}
s_1 s_3=s_2^2-2\cosh\left(\frac{\kappa \pi}{2}\right)s_2+1\,.
\end{gather}
The first solution is $S_{\ka}=C_{\ka}$.
If we take this solution for all values
of $\kappa\ne 0$ and combine it with the $s_0=-1$ solution of $\kappa=0$,
we get $S=C$, which is the star-algebra identity~(\ref{Identity}).
The second solution is actually a two-parameter family of solutions.
We parameterize the solutions using the two invariants
\begin{equation}
\label{uvEq}
\begin{split}
u&\equiv \frac{\tr S_{\ka}}{2}=\frac{s_1+s_3}2\,, \\
v&\equiv -\det S_{\ka}=2\cosh(\frac{\kappa \pi}{2})s_2-1\,,
\end{split}
\end{equation}
with the inverse relations
\begin{equation}
\begin{split}
\label{sFromuv}
s_2&=\frac{v+1}{2\cosh(\frac{\kappa \pi}{2})}\,,\\
s_{1,3}&=u\pm i\sqrt{\left(\frac{v+1}{2\cosh(\frac{\kappa \pi}{2})}\right)^2
     -v-u^2}\,,
\end{split}
\end{equation}
where $s_1$ gets the plus sign and $s_3$ gets the minus sign,
or vice versa.
Every $s_{1,2,3}$ that comes from a choice of $u,v$ is a solution
of the projection equation,
but not all
solutions are legitimate. Being a Bogoliubov transformation we have
the restriction (\ref{BogConst}) on the matrix $S_{\ka}$.
We shall allow for singular transformations,
\begin{equation}
\label{Bog2}
\lambda_{S_{\ka}^\dagger S_{\ka}}=(s_2\pm|s_1|)^2\leq 1\,,
\end{equation}
as we were forced to do in the $\kappa=0$ case.

The BPZ reality condition implies that $u,v$ are real
as well as the square root in (\ref{sFromuv}).
The condition for the solution to be twist invariant is
\begin{equation}
\label{TwistInv}
[S,C]=0 \Rightarrow s_1=s_3\,.
\end{equation}
When combined with the BPZ reality condition~(\ref{SReal}), twist
invariance enforces the reality of $S_{\ka}$.
In the $u,v$ language this condition reads
\begin{equation}
u^2=\left(\frac{v+1}{2\cosh(\frac{\kappa \pi}{2})}
     \right)^2-v\,.
\end{equation}
We will show that this solution contains the generalized
butterfly projectors, including the sliver and the butterfly.

The normalization requirement (\ref{Bog2}) and
the reality condition on $S_{\ka}$ (\ref{SReal}) restricts $u,v$ to the region
\begin{equation}
\label{uvRange}
\begin{gathered}
v\geq -1\,,\\
\left(\frac{v+1}{2\cosh(\frac{\kappa \pi}{2})}\right)^2-v\geq u^2\,.
\end{gathered}
\end{equation}
These equations determine the
allowed range for the solutions as a function of $\kappa$
as shown in figure \ref{fig:phase}.
To build a projector one has to choose an allowed value for $(u,v)$
for every value of $\kappa>0$ as illustrated in figure \ref{fig:3Dphase}.
This prescription allows for a large class of projectors.\suppressfloats[t]
\FIGURE[t]{
\label{fig:phase}
\centerline{\begin{picture}(0,0)%
\epsfig{file=Phase.pstex}%
\end{picture}%
\setlength{\unitlength}{3947sp}%
\begingroup\makeatletter\ifx\SetFigFont\undefined%
\gdef\SetFigFont#1#2#3#4#5{%
  \reset@font\fontsize{#1}{#2pt}%
  \fontfamily{#3}\fontseries{#4}\fontshape{#5}%
  \selectfont}%
\fi\endgroup%
\begin{picture}(5424,3690)(589,-3286)
\put(601,-1636){\makebox(0,0)[lb]{\smash{\SetFigFont{12}{14.4}{\rmdefault}{\mddefault}{\updefault}-1}}}
\put(5851,-1636){\makebox(0,0)[lb]{\smash{\SetFigFont{12}{14.4}{\rmdefault}{\mddefault}{\updefault}1}}}
\put(6001,-1486){\makebox(0,0)[lb]{\smash{\SetFigFont{12}{14.4}{\rmdefault}{\mddefault}{\updefault}$u$}}}
\put(3151,-3286){\makebox(0,0)[lb]{\smash{\SetFigFont{12}{14.4}{\rmdefault}{\mddefault}{\updefault}-1}}}
\put(3226,239){\makebox(0,0)[lb]{\smash{\SetFigFont{12}{14.4}{\rmdefault}{\mddefault}{\updefault}$v$}}}
\put(3151, 89){\makebox(0,0)[lb]{\smash{\SetFigFont{12}{14.4}{\rmdefault}{\mddefault}{\updefault}1}}}
\put(3976,-436){\makebox(0,0)[lb]{\smash{\SetFigFont{12}{14.4}{\rmdefault}{\mddefault}{\updefault}$e^{-\kappa\pi}$}}}
\put(2466,-286){\makebox(0,0)[lb]{\smash{\SetFigFont{12}{14.4}{\rmdefault}{\mddefault}{\updefault}$\kappa=0$}}}
\put(3366,-1401){\makebox(0,0)[lb]{\smash{\SetFigFont{12}{14.4}{\rmdefault}{\mddefault}{\updefault}$\kappa=\infty$}}}
\end{picture}
}
\vspace{-16pt}
\caption{
A projector is a curve in the $\ka,u,v$ space.
This curve can be described as a function from $\ka>0$ to a range in
the $u,v$ plane which is defined by eq. (\ref{uvRange}).
Notice that this range is $\ka$ dependent, it is colored orange for $\ka=0.5$.
In the limit $\ka\rightarrow 0$, the allowed range is the whole triangle.
But the requirement for continuity is that for $\ka=0$ the projector should
end on the right edge or the left vertex of the triangle (colored green).
}
}
\FIGURE{
\label{fig:3Dphase}
\centerline{\epsfig{file=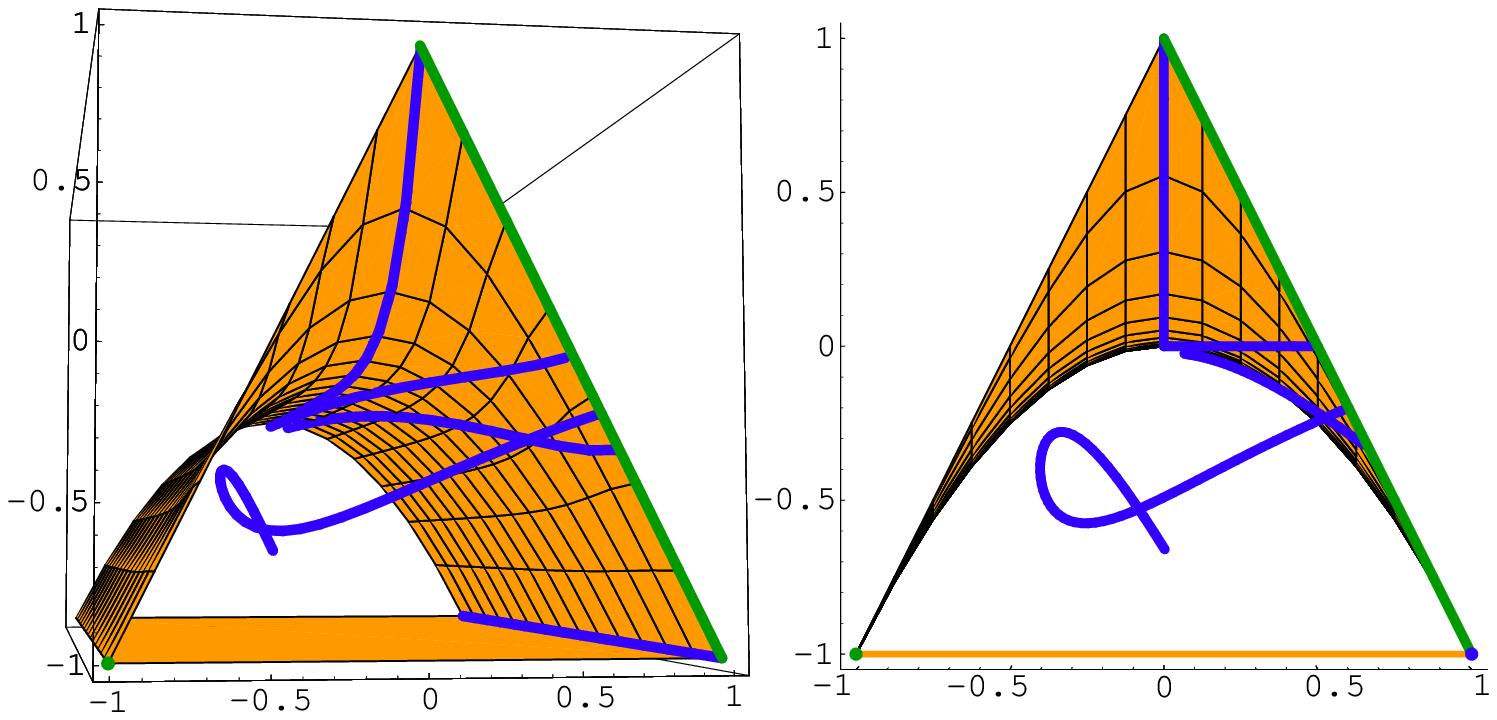}}
\vspace{-24pt}
\caption{
Various projectors in the $\ka,u,v$ space.
The depth of the figure on the left parameterizes $\ka$ with $\ka=0$
in the front and $\ka=3\approx\infty$ in the back.
The right figure is a projection of the left figure to the $u,v$ plane.
The curves on the top surface are twist invariant projectors,
they are generalized butterfly states to be described in sections
\ref{GoodBadUgly}, \ref{GBSpec}
(from top to bottom):
(i) The sliver $\alpha=0$.
(ii) The butterfly $\alpha=1$.
(iii) The state $\alpha=1.5$.
(iv) The nothing $\alpha=2$.
The additional curve represents a generic non twist invariant projector.
}
}

$u,v$ should be integrable functions of $\ka$ because
the continuous $\ka$ basis is related to the oscillator basis by integration.
Moreover, we should identify functions which differ on a
zero measure set.
Further restrictions on the functions $u,v$ would be related to
identifying which string fields are legitimate.
What is the class of legitimate string fields is still unknown.
It should be large enough to
contain D-branes \cite{Gross:2001rk,Rastelli:2001rj},
perturbative states around D-branes \cite{Hata:2001sq}, as well as closed
strings \cite{Hashimoto:2001sm}. Yet, it should not be too large
\cite{Schnabl:2002gg,Douglas:2002jm}.
Continuity of $u,v$ as a functions of $\ka$ restricts the class of
string fields. We do not know if it has anything to do with the
``correct'' choice, but we shall henceforth mention some of its consequences.
In fact without this restriction the analysis of $\ka=0$ in
section \ref{ka=0} is meaningless.

Continuity of $s_{1,2,3}$ at $\ka=0$ implies that
one of the $\ka=0$ solutions (\ref{s0Solutions}) is reached
in the $\kappa\rightarrow0$ limit.
The eigenvector $v^{(\ka=0)}$ is twist odd\cite{Rastelli:2001hh}.
Therefore, to get the relevant solution of $S_{\ka}$ we use the projector
on the twist odd eigenvalue
\begin{equation}
P_{\ka}^{(-)}=-\frac{1}{2}(C_{\ka}-\One)=\frac{1}{2}\mat{1&1\\1&1}\,,
\end{equation}
In the limit $\ka\rightarrow 0$ we get
\begin{equation}
\begin{gathered}
S_{\ka}^{(-)}=P_{\ka}^{(-)}S_{\ka}P_{\ka}^{(-)}\xrightarrow{\ka\rightarrow 0}
    \left(u+\frac{v+1}{2}\right)\cdot P_{\ka}^{(-)}=s_0\cdot P_{\ka}^{(-)}\\
\Rightarrow \begin{cases}v=1-2u & s_0=1\\
                         v=-3-2u & s_0=-1\end{cases}\,.
\end{gathered}
\end{equation}
We see that the SFT projectors (\ref{ManySolutions}) which are continuous
with respect to $\kappa$ will end either on the right segment of the figure
and have $s_0=1$, or at the point $u=v=-1$, and have $s_0=-1$.
The identity (\ref{OneSolution}) has $s_0=-1$,
since $P_{\ka}^{(-)}C_{\ka}P_{\ka}^{(-)}=-1\cdot P_{\ka}^{(-)}$.

All the projectors in (\ref{ManySolutions}) are of rank one. This follows from
the fact that a projector is given by a trajectory in the $\ka,u,v$ space.
The rank can be given by a continuous function of this trajectory,
all the trajectories are homotopic, and the previously known ones among them
are of rank one.
This observation is in accordance with the proof of \cite{Bars:2002nu} that
all Gaussian projectors apart of the identity are of rank one.
Of course both our argument, and that of \cite{Bars:2002nu} may fail if the
given projector is too singular.
Another property of rank one projectors is their factorization to
functions of the left and right part of the string.
We show that the projectors in (\ref{ManySolutions}) factorize in
\cite{FKM2}.

\subsection{The sliver, the butterfly and the nothing}
\label{GoodBadUgly}
Among our solutions we can recognize the sliver, the butterfly
and the nothing. We shall see in section~\ref{GBSpec} that in
fact all the generalized butterfly states are in $\HH_{\ka^2}$.
The spectrum of the sliver is given in \cite{Rastelli:2001hh}
\begin{equation}
\tau(\ka)=-e^{-\frac{|\ka|\pi}{2}}\,,
\end{equation}
or in our notation
\begin{equation}
\label{Sliver}
T_{\ka}=C_{\ka}S_{\ka}
    =-\exp\left(-\frac{\kappa \pi}{2}\right)\mat{1 & 0 \\ 0 & 1}\Rightarrow
S_{\ka}=\exp\left(-\frac{\kappa \pi}{2}\right)\mat{0 & 1 \\ 1 & 0}\,.
\end{equation}
Using eq. (\ref{uvEq}) we get for the sliver $u=0$, $v=\exp(-\kappa \pi)$.

For the butterfly the eigenvalues in the $C^{(+)}$ subspace are zero,
and in the $C^{(-)}$ subspace the eigenvalues 
are given by eq. (\ref{ButtEigen}), as is shown in the appendix.
We transform from the $C^{(\pm)}$ basis to our basis and get
\begin{equation}
\label{buttS}
S_{\ka}=\frac{1}{\sqrt{2}}\mat{\phantom{-}1 & 1\\-1 & 1}
    \mat{0 & 0\\0 & \frac{1}{\cosh(\frac{\kappa \pi}2)}}
    \frac{1}{\sqrt{2}}\mat{1 & -1\\1 & \phantom{-}1}
    =\frac{1}{2\cosh(\frac{\kappa \pi}2)}\mat{1 & 1 \\ 1 & 1}\,,
\end{equation}
therefore, for the butterfly $u=\frac{1}{2\cosh(\frac{\kappa \pi}2)}$, $v=0$.

The nothing is the squeezed state defined by the identity matrix,
$S=I$, therefore
\begin{equation}
\label{nothingS}
S_{\ka}=\mat{1 & 0 \\ 0 & 1}\,,
\end{equation}
that is $u=1,v=-1$.

These three states are plotted in figure \ref{fig:3Dphase}.

\section{Relations to other representations}
\label{representations}

Now that we have the form of the $\HH_{\ka^2}$ projectors in the $\K_1$
basis we would like to represent them in other forms.
We start by transforming the states in $\HH_{\ka^2}$ to the oscillator
basis.
Then we derive a procedure to find the conformal map of squeezed states.
Only surface state can pass this procedure, giving us a way to
recognize which squeezed states are surface states.
We demonstrate this procedure by some examples.

We also derive an inverse procedure for
checking if a state is in $\HH_{\ka^2}$ and if so find
the matrix $S_{\ka}$ representing it.
Finding the spectrum given $S_{\ka}$ is straightforward.
We use this procedure to show that the generalized butterfly states are
in $\HH_{\ka^2}$ and to find their spectrum.
Then we show that not all the $L_{-2m}$ projectors are in $\HH_{\ka^2}$.

We also compare the structure of the states
in $\HH_{\ka^2}$ to their form in the half-string formalism.
Transforming $S_{\ka}$ to the continuous Moyal representation is
straightforward and useful. We show that all the projectors
in $\HH_{\ka^2}$ have the same normalization in this formalism
and that the generalized butterfly states are orthogonal.
We also transform $S_{\ka}$ to the discrete Moyal representation and
show that projectors in $\HH_{\ka^2}$ satisfy the projector condition
in this formalism.

\subsection{The oscillator basis}

To calculate the matrices in the oscillator basis of states
in $\HH_{\ka^2}$ we use the orthogonality and completeness of the
$\kappa$ basis \cite{Okuyama:2002yr}.
The basis vectors are defined by
\begin{equation}
\keto{\ka}=\sum_{n=1}^\infty v_n^{(\kappa)}\keto{n}\,,
\end{equation}
where
\begin{equation}
\brao{n}=\keto{n}^T=(0,0,\ldots,0,1,0,0,\ldots)\,,
\end{equation}
with the $1$ in the $n^{th}$ position of the vector.
The coefficients
\begin{equation}
v_n^{(\kappa)}=\braketo{\ka}{n}\,,
\end{equation}
are given by the generating function
\begin{equation}
\label{GenFun}
\sum_{n=1}^\infty \frac{v_n^{(\kappa)}}{\sqrt n}z^n=
  \frac{1}{\kappa}(1-e^{-\kappa \tan^{-1}(z)})\equiv f_\kappa(z)\,.
\end{equation}
The above relation can be inverted to give
\begin{equation}
\label{GenFunInv}
v_n^{(\kappa)}=\frac{\sqrt{n}}{2 \pi i}\oint
 \frac{f_\ka(z)}{z^{n+1}} dz=
\frac{1}{\sqrt{n}}\frac{1}{2\pi i}\oint\frac{dz}{z^n}\partial_z f_\ka(z)\,,
\end{equation}
where the contour is around the origin,
and we integrated by parts to get the last expression.
The orthogonality and completeness relations in the $\ka$ basis are
\begin{gather}
\label{orthoEv}
\braketo{\kappa}{\kappa'}=\N(\kappa)\delta(\kappa-\kappa')\,,  \\
\One=\int_{-\infty}^\infty d\ka\;\frac{\keto{\ka}\brao{\ka}}{\N(\ka)}\,,
\end{gather}
where
\begin{equation}
\label{Ndef}
\N(\kappa)=\frac{2}{\kappa}\sinh\left(\frac{\kappa \pi}{2}\right)\,,
\end{equation}
is a normalization factor.

The matrix elements of a squeezed state can be 
written using a two-parameter generating function
\begin{equation}
\label{SGen}
S_{nm}=
  \frac{1}{\sqrt{nm}}\oint\frac{dz dw}{(2\pi i)^2}\frac{1}{z^{n}w^{m}}S(z,w)\,,
\end{equation}
with the inverse relation
\begin{equation}
\label{SGenInv}
S(z,w)=\sum_{m,n=1}^{\infty}\sqrt{nm}S_{nm}z^{n-1}w^{m-1}\,.
\end{equation}
The symmetry $S_{nm}=S_{mn}$ translates to the symmetry $S(z,w)=S(w,z)$.

We can now calculate the matrix elements of a state in the $\HH_{\ka^2}$
subalgebra.
That is, given $S_{\ka}$ for $\kappa>0$
the matrix elements are
\begin{multline}
\label{Sfroms}
S_{nm}=
\brao{n}S\keto{m}=\int_{-\infty}^\infty \frac{d\ka d\ka'}{\N(\ka)\N(\ka')}
\braketo{n}{\ka}\brao{\ka}S\keto{\ka'}\braketo{\ka'}{m}=\\
  \int_{0}^\infty \frac{d\ka d\ka'}{\N(\ka)\N(\ka')}
    \mat{\braketo{n}{\sm\ka} & \braketo{n}{\ka}}
    \mat{\brao{\sm\ka}S\keto{\sm\ka'} & \brao{\sm\ka}S\keto{\ka'}\\
  \phantom{\sm}\brao{\ka}S\keto{\sm\ka'} & \phantom{\sm}\brao{\ka}S\keto{\ka'}}
    \mat{\braketo{\sm\ka'}{m}\\\phantom{\sm}\braketo{\ka'}{m}}=\\
  \int_{0}^\infty \frac{d\ka}{\N(\ka)}
    \mat{\braketo{n}{\sm\ka} & \braketo{n}{\ka}}
    \mat{s_1(\ka) & s_2(\ka)\\s_2(\ka) & s_3(\ka)}
    \mat{\braketo{\sm\ka}{m}\\\phantom{\sm}\braketo{\ka}{m}}=\\
  \frac{1}{\sqrt{nm}}\oint\frac{dz dw}{(2\pi i)^2}\frac{1}{z^{n}w^{m}}
    \int_{0}^\infty \frac{d\ka}{\N(\ka)}
    \mat{\partial_z f_{-\ka}(z) & \partial_z f_\ka(z)}
    S_{\ka}
    \mat{\partial_w f_{-\ka}(w) \\ \partial_w f_\ka(w)}\,.
\end{multline}
Notice that we limited ourselves to the $\HH_{\ka^2}$ subalgebra
in passing from the second line to the third line.
In the last equality we used (\ref{GenFunInv}). By defining
\begin{align}
s_{13}(\ka)&\equiv\begin{cases}
       s_3(\ka) & \ka>0 \\
       s_1(-\ka) & \ka<0\end{cases}\,,&
s_{22}(\ka)&\equiv\begin{cases}
       s_2(\ka) & \ka>0 \\
       s_2(-\ka) & \ka<0\end{cases}\,,
\end{align}
we get
\begin{multline}
\label{Sdef}
S(z,w)=\int_{-\infty}^\infty \frac{d\ka}{\N(\ka)}
    \left(\partial_z f_\ka(z)\partial_w f_\ka(w)s_{13}(\ka) +
    \partial_z f_\ka(z)\partial_w f_{-\ka}(w)s_{22}(\ka)\right)=\\
  \int_{-\infty}^\infty \frac{d\ka}{\N(\ka)}
    \frac{e^{-\ka(\tan^{-1}(z)+\tan^{-1}(w))}s_{13}(\ka)+
   e^{\ka(\tan^{-1}(w)-\tan^{-1}(z))}s_{22}(\ka)}{(1+z^2)(1+w^2)}\,.
\end{multline}
This equation is what we were after: the oscillator matrix
elements as a function of $S_{\ka}$.

\subsection{Surface states}

The matrix elements expression (\ref{SGen}) for squeezed states has
a similar structure to that of
surface states \cite{LeClair:1989sp,LeClair:1989sj,Gaiotto:2002kf}
\begin{equation}
\label{SSurface}
S_{nm}=-\frac{1}{\sqrt{nm}}
  \oint\frac{dzdw}{(2\pi i)^2 z^{n}w^{m}}
\frac{f'(-z)f'(-w)}{(f(-z)-f(-w))^2}\,,
\end{equation}
where $f(z)$ is the conformal transformation that defines the
surface state.
This raises the question which squeezed states
are surface states, and what is their conformal transformation $f(z)$.
Squeezed states are surface states if and only if $\exists f(z)$
such that
\begin{equation}
S(z,w)\approx -\frac{f'(-z)f'(-w)}{(f(-z)-f(-w))^2}\,.
\end{equation}
By $\approx$ we mean ``equal up to poles'',
since poles do not contribute to the contour integrals in eq.~(\ref{SSurface}).
However $S(z,w)$ is regular near the origin, and
\begin{equation}
\frac{f'(-z)f'(-w)}{(f(-z)-f(-w))^2}=\frac{1}{(z-w)^2}+
  \mbox{regular terms}\,.
\end{equation}
Therefore, the condition is that there exists $f(z)$ such that
\begin{equation}
\label{surfaceEq}
S(z,w)=-\frac{f'(-z)f'(-w)}{(f(-z)-f(-w))^2}+
\frac{1}{(z-w)^2}\,.
\end{equation}
If this is the case, then in particular
\begin{equation}
S(z,0)=-\frac{f'(-z)}{f(-z)^2}+\frac{1}{z^2}\,,
\end{equation}
where we used $f(0)=0, f'(0)=1$ which is possible due to $SL(2,\mathbb C)$
invariance.
The solution of this equation
gives us a candidate for $f(z)$
\begin{equation}
\label{surfaceSol}
f^{\text{c}}(z)=\frac{z}{1-z \int_0^{-z} S(\tilde z,0)d\tilde z}.
\end{equation}
Given $S(z,w)$, one can solve eq. (\ref{surfaceSol}) to get
$f^{\text{c}}(z)$,
then substitute the solution in eq. (\ref{surfaceEq}) and check
if it reproduces $S(z,w)$. 
A squeezed state is a surface state if and only if
$f^{\text{c}}(z)$ reproduces $S(z,w)$.
We now turn to some examples.

\subsubsection{Reconstructing the butterfly}

To evaluate the generating function (\ref{Sdef}) of the
butterfly (\ref{buttS}) we use the relation
\begin{equation}
\int_{-\infty}^\infty d\kappa
\frac
{e^{c\kappa}}{\frac{2}{\kappa}\sinh\left(\frac{\kappa \pi}{2}\right)
  2\cosh(\frac{\kappa \pi}2)}=\frac{1}{4\cos({\frac{c}{2}})^2}\,.
\end{equation}
Replacing $c$ by $-\tan^{-1}(z)\pm\tan^{-1}(w)$ we obtain
\begin{equation}
\label{buttSgen}
S(z,w)=\frac{w^2+z^2-\frac{w^2+z^2+2w^2z^2}{\sqrt{(1+w^2)(1+z^2)}}}
  {(w^2-z^2)^2}\,.
\end{equation}
By eq. (\ref{surfaceSol}), we get
\begin{equation}
\label{fOfButt}
f(z)=\frac{z}{\sqrt{1+z^2}}\,,
\end{equation}
which is the correct expression for the butterfly.
Substituting this expression into (\ref{surfaceEq}) reproduces
eq. (\ref{buttSgen}), as it should.

\subsubsection{The duals of the nothing and of the identity states}

We demonstrated in \ref{GoodBadUgly} that the nothing state has
$u=1,v=-1$, that is $S_{\ka}=\One$.
The dual of the nothing has $S_{\ka}=-\One$, and is the
mirror image of the nothing in the $u,v$ plane, $u=v=-1$.
Both projectors live at the boundary of figure \ref{fig:phase},
and saturate the inequality (\ref{surfaceEq})
for all $\kappa$, and are therefore very singular.
It can be shown that
the mirror images of the other butterfly states do not
correspond to continuous projectors.

Equation (\ref{Sdef}) now gives
\begin{equation}
\label{Sofu=v}
S(z,w)=-\frac{1}{(1-wz)^2}\,.
\end{equation}
Now we use eq.
(\ref{surfaceSol}) to find the candidate $f(z)$
\begin{equation}
\label{u=v=-1}
f^{\text{c}}(z)=\frac{z}{1-z^2}\,.
\end{equation}
We recognize $f^{\text{c}}(z)$ as the conformal transformation of
the identity state, which is represented by $C$, rather by $-I$.
Indeed, when we substitute (\ref{u=v=-1}) back into
(\ref{surfaceEq}) we get
\begin{equation}
S^{\text{Id}}(z,w)=-\frac{1}{(1+wz)^2}\,,
\end{equation}
instead of (\ref{Sofu=v}).
This proves that the dual of the nothing state is not a surface state.

The dual of the identity $S_{\ka}=-C_{\ka}$ is not a projector.
It has
\begin{equation}
f^{\text{c}}(z)=\frac{z}{1+z^2}\,,
\end{equation}
which is the conformal map of the nothing, meaning that it is also
not a surface state.

\subsubsection{A non-orthogonal projector}

Here we want to give an example of another surface state projector.
This projector has $v=0,\ u=\frac{1}{2}$ for all $\ka$.
Notice that it does not respect the reality condition (\ref{SReal})
meaning that it is a non-orthogonal projector as is discussed in
section (\ref{HalfString}).
In the $\K_1$ basis the projector is
\begin{align}
s_{13}(\ka)&=\frac{1}{2}\left(1\pm\tanh\left(\frac{\ka\pi}{2}\right)\right)\,,\\
s_{22}(\ka)&=\frac{1}{2\cosh\left(\frac{\ka\pi}{2}\right)}\,.
\end{align}
Note that we are describing two projectors $(\pm)$.
Non-orthogonal projectors come in pairs, because they are asymmetric.
Therefore, if we find one projector, its twisted partner will give another.
The conformal transformations of the projectors are
\begin{equation}
f(z)=\pm 1+\frac{z\mp 1}{\sqrt{1+z^2}}\,.
\end{equation}
We can alway build two orthogonal projectors from a non-orthogonal pair
by star multiplying them in different orders.
The current states are a gluing of
the generalized butterfly $(\alpha=\frac{2}{3})$ on one side
with the nothing $(\alpha=2)$ on the other~\cite{FKM2}.

\subsection{The inverse transformation}

In the previous subsections we constructed the matrix representation
and the conformal map representation for states in the $\HH_{\ka^2}$
subalgebra.
It is natural to ask the opposite question.
Does a given state belongs to $\HH_{\ka^2}$, and if so,
what is its form in the $\K_1$ basis.
Given its form in the $\K_1$ basis we can immediately check if it is
a projector (\ref{ManySolutions}).

In the appendix we do it explicitly for the butterfly.
Here we will find a general prescription for all states
by examining when is it possible to invert the relations of the
previous subsection.

Given $f(z)$ or $S_{nm}$ we can use eq.~(\ref{surfaceEq}), or
eq.~(\ref{SGenInv}) to define $S(z,w)$.
The question is, what are the conditions
for the existence of $s_{13}(\kappa)$, $s_{22}(\kappa)$ which reproduce
$S(z,w)$ via eq. (\ref{Sdef}), and what are
those $s_{13}(\kappa)$, $s_{22}(\kappa)$.

Inspecting eq. (\ref{Sdef}), we see that $S(z,w)(1+z^2)(1+w^2)$
cannot be an arbitrary function of $z,w$ for states in the $\HH_{\ka^2}$
subalgebra, but should rather be a sum of two terms
\begin{equation}
\label{split}
S(z,w)(1+z^2)(1+w^2)=F_1(\xi)+F_2(\zeta)\,,
\end{equation}
where we have defined
\begin{equation}
\begin{split}
\label{XiZeta}
\xi&=i(\tan^{-1}(z)+\tan^{-1}(w))\,,\\
\zeta&=i(\tan^{-1}(z)-\tan^{-1}(w))\,.
\end{split}
\end{equation}
From eq. (\ref{Sdef}) we see that $F_1(\xi)$ is the Fourier transform
of $s_{13}(\kappa)/\N(\kappa)$, while
$F_2(\zeta)$ is the Fourier transform of $s_{22}(\kappa)/\N(\kappa)$.
We conclude that this split to a sum of functions is a necessary
condition for the state to be in $\HH_{\ka^2}$ due to
the form of eq. (\ref{Sdef}), and a sufficient condition since the
Fourier transform is invertible.

Suppose now that $F_1$ and $F_2$ are given.
The inverse Fourier transform reads
\begin{equation}
\label{invs}
\begin{split}
s_{13}(\kappa)&=\frac{\N(\kappa)}{2\pi}\int_{-\infty}^{\infty}
  e^{-i \kappa \xi}F_1(\xi)\,d\xi\,, \\
s_{22}(\kappa)&=\frac{\N(\kappa)}{2\pi}\int_{-\infty}^{\infty}
  e^{-i \kappa \zeta}F_2(\zeta)\,d\zeta\,.
\end{split}
\end{equation}
These are the desired expressions for inverting the transformation.
They apply only to states in $\HH_{\ka^2}$, meaning,
states of the form~(\ref{split}).
We can use eq.~(\ref{invs},\ref{ManySolutions}) to check if a given
state is a projector, as is illustrated below.

\subsubsection{The spectrum of the generalized butterfly states}
\label{GBSpec}
The generalized butterfly states are a one-parameter family of
projectors \cite{Gaiotto:2002kf}.
They are defined by the maps
\begin{equation}
\label{GenButt}
f(z)=\frac{1}{\alpha}\sin(\alpha \tan^{-1}(z))\,,
\end{equation}
where $0\leq \alpha \leq 2$.
Special cases include the sliver $\alpha=0$, for which
\begin{equation}
f(z)=\tan^{-1}(z)\,,
\end{equation}
the canonical butterfly $\alpha=1$ (\ref{fOfButt}), and
the nothing state $\alpha=2$ with
\begin{equation}
f(z)=\frac{z}{1+z^2}\,.
\end{equation}

We substitute the map (\ref{GenButt}) in eq.(\ref{surfaceEq}),
and notice that indeed it splits according to eq. (\ref{split}),
with
\begin{equation}
\begin{split}
F_1(\xi)&=\frac{\alpha^2}{4\cosh \left( \frac{\alpha \xi}{2}\right)^2}\,, \\
F_2(\zeta)&=\frac{\alpha^2}{4\sinh 
     \left( \frac{\alpha \zeta}{2}\right)^2}-\frac{1}{\sinh(\zeta)^2}\,,
\end{split}
\end{equation}
which means that the generalized butterfly states are in $\HH_{\ka^2}$.
Using eq. (\ref{invs}) we get
\begin{equation}
\label{gens}
\begin{split}
s_1&=s_3=s_{13}=\frac{\sinh\left(\frac{\ka\pi}{2}\right)}
    {\sinh\left(\frac{\ka\pi}{\alpha}\right)}\,,\\
s_2&=s_{22}=\cosh\left(\frac{\ka\pi}{2}\right) -
  \coth\left(\frac{\ka\pi}{\alpha}\right)\sinh\left(\frac{\ka\pi}{2}\right)\,.
\end{split}
\end{equation}
To evaluate the integrals, we closed the integration contour with
a semi-circle around the lower half plane, picking up an infinite 
number of residues along the lower half of the imaginary axis, 
which summed up to the given results.

A simple consistency check shows that these matrix elements satisfy
eq.~(\ref{ManySolutions}), which verifies that the generalized butterfly
states are indeed projectors.

\subsubsection{The $L_{-2m}$ projectors}

Another class of projectors with a simple Virasoro representation was
introduced in~\cite{Gaiotto:2002kf}. These projectors are surface states
defined by the conformal maps
\begin{equation}
\label{f-m}
f_m(z)=\frac{z}{(1-(-z^2)^m)^{1/2m}}\,,
\end{equation}
with associated states
\begin{equation}
\label{L-2m}
\ket{P_{2m}}=\exp\left(\frac{(-1)^m}{2m}L_{-2m}\right)\ket{0}\,.
\end{equation}
In order to check if a given surface state is in $\HH_{\ka^2}$,
we found it useful to use a condition which is equivalent
to eq.~(\ref{split}) for surface states
\begin{equation}
\Diamond\Box\log\left(\frac{f(-z)-f(-w)}{z-w}\right)=0\,,
\end{equation}
where $\Box=\frac{d^2}{d\xi^2}-\frac{d^2}{d\zeta^2}$ is the
two dimensional d'Alembertian and
$\Diamond=\frac{d^2}{d\xi\,d\zeta}$
is the light-cone d'Alembertian, or vice versa.
The relation between $z,w$ and $\xi,\zeta$ is given by
eq.~(\ref{XiZeta}).
This condition holds, as expected, for the case of the butterfly $m=1$.
A direct check shows that this equation does not hold for the
next 100 cases and therefore that these projectors are not in
the $\HH_{\ka^2}$ subalgebra.
We expect that higher $m$ projectors will not fulfill
this condition either.

\subsection{Half-string representation}
\label{HalfString}

In this subsection we discuss how states in the $\HH_{\ka^2}$ subalgebra
look in the half-string formalism
\cite{Bordes:1991xh,Gross:2001rk,Rastelli:2001rj,Furuuchi:2001df}.
Finding rank one projectors in half-string formalism
is extremely simple.
Every normalized string wave functional $\Psi_P [x(\sigma)]$ that has a
factorized form
\begin{equation}
\Psi_P [x(\sigma)]=\chi_1[l(\sigma)]\chi_2[r(\sigma)]\,,
\end{equation}
where $l(\sigma)\sim x(\sigma)$ and $r(\sigma)\sim x(\pi-\sigma)$
are the left and right sides of the 
string~\footnote{There are different approaches in defining half
string coordinates according to whether or not the midpoint $x(\pi/2)$
is subtracted~\cite{Gross:2001rk,Rastelli:2001rj}. This
determines the half string boundary conditions and is related to
the associativity
anomaly~\cite{Bars:2002bt} of the star product.},
corresponds to a rank one projector.
The string field reality condition~(\ref{RealState}) for a string
functional reads
\begin{equation}
\label{HSReal}
\Psi_P^*[x(\sigma)]=\Psi_P[x(\pi-\sigma)]
\Rightarrow\chi_2[r(\sigma)]=\chi_1^*[r(\sigma)]\,.
\end{equation}
String fields can be regarded as matrices over the space of half-strings.
The BPZ reality condition implies that these matrices are Hermitian.
A Hermitian projector is denoted ``orthogonal projector'' since its
zero and one eigenvalue subspaces are orthogonal.

In~\cite{Gross:2001rk,Rastelli:2001rj} projectors
with real Gaussian functionals were considered.
We consider also complex Gaussians because they appear in $\HH_{\ka^2}$
\begin{equation}
\label{ML}
\begin{split}
\Psi[X(\sigma)]  
&= \exp\left(-\frac{1}{2}l_{2k-1} M_{2k-1,2j-1}l_{2j-1}\right)
  \exp\left(-\frac{1}{2}r_{2k-1} M^*_{2k-1,2j-1}r_{2j-1}\right) \\
&=\exp\left(-\frac{1}{2}x_n L_{nm}x_m\right) \, ,
\end{split}
\end{equation}
where $l_{2k+1},r_{2j+1}$ and $x_n$ are the Fourier modes of $l(\sigma),r(\sigma)$ and $x(\sigma)$.
The form of the Gaussians in the half-string formalism was restricted
by the string field reality condition.
For the full-string matrix $L$ this condition reads
\begin{equation}
L^*=CLC\,.
\end{equation}
Solving eq.~(\ref{ML}) for $L$ gives
\begin{equation}
\begin{split}
L_{2n-1,2m-1} &=2 \, {\text Re}\, M_{2n-1,2m-1} \,,\\
L_{2n,2m} &=2 \,(T \, {\text Re}\, M \, T^T)_{2n,2m}\,, \\
L_{2n,2m-1} =L_{2m-1,2n} &=2i\, (T\, {\text Im}\, M)_{2n,2m-1}   \, ,
\end{split}
\end{equation}
where the matrix $T$ is defined by
\begin{equation}
\begin{split}
\label{Tdef}
T_{2n,2m-1}&=\,\frac{4}{\pi}\int _{0}^{\frac{\pi }{2}}
    \cos (2n\sigma)\cos (\left( 2m - 1 \right) \sigma )\,d\sigma= \\
&=\,
\frac{2(-1)^{m+n+1}}{\pi}\left(\frac{1}{2m-1+2n}+\frac{1}{2m-1-2n}\right)\,.
\end{split}
\end{equation}
The relation between Gaussian functionals and squeezed states is given
by 
\begin{equation}
\label{SofELE}
S=\frac{1-ELE}{1+ELE}\,,
\end{equation}
where the diagonal matrix $E$ is defined as in~\cite{Gross:2001rk}.

The case of a twist invariant projector is given by a real matrix $M$,
which corresponds to a real, block diagonal $S_{nm}$. 
The form of $S_{nm}$ restricts the Taylor expansion of $S(z,w)$ 
to monomials $z^k w^l$ with real coefficients,
where $k$ and $l$ are both odd or both even.
Inspection of the symmetry properties of the integrand in
eq.~(\ref{Sdef}) shows that 
this requirement is satisfied only if $s_3(\ka)=s_1(\ka)$ which is the
twist invariance condition (\ref{TwistInv}).

The case of a general orthogonal projector, i.e. a BPZ real one,
eq.~(\ref{HSReal}), is given by a complex matrix $M$.
In this case $S_{nm}$ can have imaginary entries in the odd--even blocks.
The analog condition for states in $\HH_{\ka^2}$ is $s_3(\ka)=s_1^*(\ka)$.
Non-orthogonal, non-real projectors can have general $s_1,s_2,s_3$
and general $S_{nm}$ matrices limited only by the projection condition.

\subsection{Moyal representation}
\label{Moyal}

The star algebra can be represented as an infinite tensor
product of two dimensional Moyal spaces. Two such representation were found.
The discrete one by Bars \cite{Bars:2001ag}
(see also Bars and Matsuo \cite{Bars:2002nu}),
and the continuous one by Douglas, Liu, Moore and Zwiebach 
\cite{Douglas:2002jm} (DLMZ below).
Both are intimately related to the $\K_1$ basis.

We shall start with the continuous Moyal representation
and then turn to the discrete one.

\subsubsection{Continuous Moyal representation}
\label{contMoyal}

In this representation for each $\kappa>0$ there is a pair of noncommutative
coordinates $x_\kappa, y_\kappa$ with $*$ commutation relation
\begin{equation}
\label{contMoyalBra}
[x_\kappa,y_{\kappa'}]_*=\theta(\kappa)\delta(\kappa-\kappa')\,,
\end{equation}
where
\begin{equation}
\theta(\kappa)=2\tanh\left(\frac{\ka\pi}{4}\right)\,,
\end{equation}
is the noncommutativity parameter.
These two coordinates are related to creation and annihilation operators by
\begin{equation}
x_\kappa=\frac{i}{\sqrt{2}}(e_\kappa-e_\kappa^\dagger)\,,\qquad
y_\kappa=\frac{i}{\sqrt{2}}(o_\kappa-o_\kappa^\dagger)\,,
\end{equation}
and $e_\kappa^\dagger, (o_\kappa^\dagger)$ are related to the twist even (odd)
creation operators by
\begin{equation}
\label{eoOfa}
e_\ka^\dagger=\sqrt{2}\sum_{n=1}^\infty
    \frac{v_{2n}^{(\ka)}}{\sqrt\N}a_{2n}^\dagger\,,\qquad
o_\ka^\dagger=-i \sqrt{2}\sum_{n=1}^\infty
    \frac{v_{2n-1}^{(\kappa)}}{\sqrt\N}a_{2n-1}^\dagger\,,
\end{equation}
where $v_n^{(\kappa)}, \N$ are given by eq.~(\ref{GenFunInv},\ref{Ndef}).
Note that
$v_{n}(\kappa)$ in DLMZ is equal to $\frac{v_{n}^{(\kappa)}}{\sqrt\N}$ here.

From this definition it is straightforward to transform the matrix
$S_{\ka}$ of the $\HH_{\ka^2}$ subalgebra
to the $e_\kappa^\dagger,o_\kappa^\dagger$ basis
\begin{equation}
\label{tildeS}
\tilde S_\kappa=\mat{s_{ee} & i s_{eo} \\i s_{eo} & -s_{oo}}\,,
\end{equation}
where
\begin{equation}
\mat{s_{ee} & s_{eo} \\ s_{eo} & s_{oo}}=
\frac{1}{\sqrt{2}}\mat{1 & -1\\1 & \phantom{-}1}
    \mat{s_1 & s_2\\s_2 & s_3}
    \frac{1}{\sqrt{2}}\mat{\phantom{-}1 & 1\\-1 & 1}\,.
\end{equation}
Note the $i$ factor in the definition of $o_\kappa^\dagger$ in
eq.~(\ref{eoOfa}), which was put there to make the BPZ conjugation
relations of $o_\kappa$ and $e_\kappa$ the same
\begin{equation}
BPZ(e_\kappa)=-e_\kappa^\dagger\,,\qquad
BPZ(o_\kappa)=-o_\kappa^\dagger\,.
\end{equation}
This is the source of the $i$ factor in eq.~(\ref{tildeS}),
which when combined with the reality condition (\ref{SReal}),
translates to the reality of the matrix $\tilde S_\kappa$.
The projection condition~(\ref{ManySolutions}) now takes the form
\begin{equation}
\label{tilSproj}
1+\det(\tilde S_\ka)+\cosh\left(\frac{\ka\pi}{2}\right)\tr(\tilde S_\ka)=0\,.
\end{equation}

The matrix $\tilde S_\kappa$ corresponds in coordinate space to the matrix
\begin{equation}
\label{tilLtilS}
\tilde L_\kappa=\frac{1-\tilde S_\kappa}{1+\tilde S_\kappa}\,,
\end{equation}
where the wave function in $\vec X_\kappa \equiv (x_\kappa,y_\kappa)$
is proportional to the Gaussian
\begin{equation}
\exp\left(-\frac{1}{2}\int_0^\infty
    \vec X_\kappa \tilde L_\kappa \vec X_\kappa d\kappa\right)\,.
\end{equation}
The generalized butterfly~(\ref{gens}) in this representation is
\begin{equation}
\tilde L_\kappa=\coth\left(\frac{\kappa \pi}{4}\right)
    \mat{\tanh(\frac{\kappa \pi(2-\alpha)}{4\alpha}) & 0 \\
         0  & \coth(\frac{\kappa \pi(2-\alpha)}{4\alpha}) }\,.
\end{equation}

We can think of these solutions as noncommutative solitons
\cite{Gopakumar:2000zd,Chen:2002jd}.
In the sliver limit $\alpha \rightarrow 0$ we get
\begin{equation}
\tilde L_\kappa=\coth\left(\frac{\kappa\pi}{4}\right)\mat{1 & 0 \\ 0 & 1 }\,,
\end{equation}
in agreement with \cite{Chen:2002jd}.
The generalized butterfly states can now be identified as
non-radial generalizations of the sliver,
with an opposite rescaling of $x_\kappa,y_\kappa$,
which is $\alpha$ and $\kappa$ dependent.
Rescaling a projector gives again a projector because
it does not change the brackets (\ref{contMoyalBra}).
All our twist invariant solutions (\ref{TwistInv}) are of this type.

Plugging the inverse of eq.~(\ref{tilLtilS}) into the projector
condition~(\ref{tilSproj}), one gets the elegant condition
\begin{equation}
\label{moyalCond}
\det(\tilde L_\ka)=\coth^2\left(\frac{\ka\pi}{4}\right)=
    \frac{4}{\theta^2(\ka)}\,.
\end{equation}
We now show that the projectors of $\HH_{\ka^2}$ can be interpreted as
a change of coordinates
which does not change the brackets~(\ref{contMoyalBra}).
Under a change of coordinates $\vec X \rightarrow U\vec X$, we get
$\tilde L_\kappa \rightarrow {U^{-1}}^T\tilde L_\kappa U^{-1}$.
The reality of the change of coordinates,
together with the invariance of the brackets means that $U\in SP(2,\mathbb R)$.
The reality and positivity of $\tilde L_\kappa$ together with
the invariance of $\det(\tilde L_\kappa)$ means that $U\in SL(2,\mathbb R)$.
But these two groups are the same.
Note that $SP(2,\mathbb R)$ has three generators, while $\tilde L_\kappa$
has only two degrees of freedom.
For each $\tilde L_\kappa$ there exists a generator of $SP(2,\mathbb R)$,
which acts trivially on it.

$\tilde L_\kappa$ can be parametrized as
\begin{equation}
\tilde L_\kappa=
\mat{l_1^{(\kappa)} & l_2^{(\kappa)} \\ l_2^{(\kappa)} & l_3^{(\kappa)}}=
\coth\left(\frac{\kappa \pi}{4}\right)
\mat{r(\kappa)^{-1}\cosh(\phi(\kappa)) & \sinh(\phi(\kappa)) \\
  \sinh(\phi(\kappa)) & r(\kappa)\cosh(\phi(\kappa))}\,.
\end{equation}
A continuous solution corresponds to a continuous choice of $r(\kappa),\phi(\kappa)$.
Continuity at $\kappa=0$ translates into the requirement that
as $\kappa$ goes to zero the term
$\coth(\frac{\kappa \pi}{4})r(\kappa)\cosh(\phi(\kappa))$ either
diverges ($s_0=1$), or goes to
zero ($s_0=-1$).

This can be generalized to a change of coordinates,
for which $\HH_{\ka^2}$ is not invariant, that mixes
$x_\kappa,y_\kappa$ for different values of $\kappa$ without changing the
brackets.
Some kind of continuity should be imposed here as well, in order to define
this group of transformations.
With the notion of continuity made clear,
this more general symmetry can be used for the production
of general Gaussian projectors.
Presumably all rank one Gaussian projectors (such as the $L_{-2n}$, which
are not in $\HH_{\ka^2}$) can be represented this way.
We shall return to this point in section \ref{discMoyal}.
When non-Gaussian projectors are included, they are connected by yet another
gauge group \cite{Gopakumar:2000zd}.
The most general gauge symmetry, which connects
different rank-one projectors, should be generated by these two groups.

To normalize the $\HH_{\ka^2}$ projectors in the DLMZ conventions we calculate
\begin{multline}
\int Dx(\kappa)Dy(\kappa)\exp\left(-\int_0^\infty d\kappa
    \vec X_\ka \tilde L_\kappa \vec X_\ka\right)=\\
\exp\left(-\frac{1}{4}\frac{\log L}{2\pi}\int_0^\infty d\kappa
    \log \det(\tilde L_\kappa)\right)=
    \exp\left(-\frac{\log L}{8}\right)\,,
\end{multline}
where $L$ is the level cutoff, using eq.~(\ref{moyalCond}). 
The normalized state is then given by
\begin{equation}
\label{ProjNorm}
\exp\left(\frac{\log L}{16}\right)
    \exp\left(-\frac{1}{2}\int_0^\infty
        \vec X_\ka \tilde L_\ka \vec X_\ka d\ka\right)\,.
\end{equation}
The normalization factor is infinite, since we should take the cutoff
$L\rightarrow\infty$.
It is usually claimed that this would be compensated by the ghost factor, although
it may be not that simple \cite{Erler:2002nr,Okuyama:2002yr}.

Using this normalization we now show that,
at least in the matter sector, different Gaussian projectors are orthogonal,
and thus can be used to construct multi $D$-brane solutions with no need
to include non-Gaussian projectors \cite{Gross:2001rk,Rastelli:2001rj}.
This stems from the fact that for rank one projectors $\braket{\Psi_1}{\Psi_2}=0$
is equivalent to $\Psi_1 * \Psi_2=0$, as can be easily seen in the half string
representation.

The inner product of two generalized butterfly states
parametrized by $a_i\equiv\frac{2-\alpha_i}{\alpha_i}$ is
\begin{multline}
\label{ButtOrth}
\exp\left(\frac{\log L}{8}\right)
\int Dx(\kappa)Dy(\kappa)\exp\left(-\int_0^\infty d\kappa\frac{1}{2}
\vec X_\ka (\tilde L_\ka^{a_1}+\tilde L_\ka^{a_2})\vec X\ka\right)=\\
\exp\left(\frac{\log L}{8}-\frac{1}{4}\frac{\log L}{2\pi}\int_0^\infty d\kappa
 \log\det\frac{\tilde L_\kappa^{a_1}+\tilde L_\kappa^{a_2}}{2}\right)=\\
\exp\left(-\frac{\log L}{8\pi}\int_0^\infty d\kappa
 \log\frac{2+\tanh(a_1\ka)\coth(a_2\ka)+
 \tanh(a_2\ka)\coth(a_1\ka)}{4}\right)=\\
\exp\left(-\frac{\log L}{48}\frac{(a-b)^2}{ab(a+b)}\right)\,.
\end{multline}
For $a\neq b$ the inner product goes to zero when the cutoff $L$
is sent to infinity.
This orthogonality of two Gaussians is only possible
due to the fact that our space of integration is infinite dimensional.
Indeed for any finite $L$ the overlap is nonzero,
and thus the orthogonality is only approximate in a cut-off theory.

The fact that two different butterflies are orthogonal is not in
conflict with the known fact that the inner product of two surface states
equals one \cite{LeClair:1989sj}.
This is because we are considering only the matter sector.
The orthogonality calculation is meaningful if we take two different
butterflies in the matter sector with the same structure in the ghost sector.
Thus at least one of them would not be a surface state.
One should be aware that such ``unbalanced states''
may be problematic \cite{Schnabl:2002gg}.

\subsubsection{Discrete Moyal representation}
\label{discMoyal}

The basic $*$ commutation relation in this representation
\cite{Bars:2001ag,Bars:2002nu} is
\begin{equation}
[x_{2n},p_{2m}]_*=i \theta \delta_{m,n}\,,
\end{equation}
where $\theta$ is an arbitrary parameter, $x_{2n}$ are the even modes of the string, and $p_{2n}$
are linear combinations of the odd momenta of the string
\begin{equation}
\label{PePo}
p_{2m}=\frac{\theta}{2}\sum_{n=1}^\infty p_{2n-1}R_{2n-1,2m}\,,\qquad
p_{2n-1}=\frac{2}{\theta}\sum_{n=1}^\infty p_{2m}T_{2m,2n-1}\,.
\end{equation}
The matrix $R$ is the inverse of the matrix $T$~(\ref{Tdef}).

$\HH_{\ka^2}$ states should be Gaussians in the variables $x_{2n},p_{2n}$.
To see their form in this formulation we can use the transformation to the
oscillator basis eq.~(\ref{Sfroms}), and then use the inverse of eq.~(\ref{SofELE}),
Fourier transform the odd modes, and use the $T$ matrix to get the final form.
The other option would be to use the form of $\HH_{\ka^2}$ states in the
continuous Moyal representation,
and then transform them to the discrete one \cite{Douglas:2002jm}.
We shall use this method in order to avoid the technical problems of
inverting infinite dimensional matrices. 
We shall need eq.~(4.11,4.12) of DLMZ, which
we present here in our notations
\begin{equation}
\begin{split}
\label{xKappa2n}
x_\kappa=&\phantom{-}\sqrt{2}\sum_{n=1}^\infty \frac{v_{2n}^{(\kappa)}\sqrt{2n}}
    {\sqrt{\N(\kappa)}}x_{2n}\,, \\
y_\kappa=&-\sqrt{2}\sum_{n=1}^\infty \frac{v_{2n-1}^{(\kappa)}}
    {\sqrt{\N(\kappa)}\sqrt{2n-1}}p_{2n-1}\,.
\end{split}
\end{equation}

Squeezed states in the discrete Moyal representation read
\begin{equation}
\exp(-\xi_i M_{ij} \xi_j)\,,
\end{equation}
where $M$ is an infinite matrix and $\xi=(x_2,x_4,...,p_2,p_4,...)$.
Our task is to find the transformation from $\tilde L_\kappa$ to $M$ using
\begin{equation}
\label{MofL}
\exp(-\xi M \xi)=\exp\left(
-\frac{1}{2}\int_0^\infty \vec X_\ka \tilde L_\ka \vec X_\ka d\ka\right)\,.
\end{equation}
We write the matrix $M$ as
\begin{equation}
M=\mat{M_1 & M_2 \\ M_2^T & M_3}\,,
\end{equation}
where $M_1,M_2,M_3$ are infinite matrices, and $M_1,M_3$ are symmetric.
Now we substitute eq.~(\ref{xKappa2n},\ref{PePo}) into eq.~(\ref{MofL}),
and compare coefficients to get
\begin{equation}
\begin{split}
\label{m1m2m3}
(M_1)_{2m,2n}=&
\int_0^\infty d\kappa\frac{v_{2m}^{(\kappa)} v_{2n}^{(\kappa)}\sqrt{(2m)(2n)}}
{\N(\kappa)}l_1^{(\kappa)}\,,\\
(M_2)_{2m,2n}=&
-\int_0^\infty d\kappa
 \frac{v_{2m}^{(\kappa)} \sqrt{2m}}{\N(\kappa)}l_2^{(\kappa)}\frac{2}{\theta}
   \sum_{k=1}^\infty T_{2n,2k-1}\frac{v_{2k-1}^{(\kappa)}}{\sqrt{2k-1}}= \\
=&
-\frac{2}{\theta}\int_0^\infty d\kappa
 \frac{v_{2m}^{(\ka)}v_{2n}^{(\ka)} \sqrt{2m}}{\N(\ka)\sqrt{2n}}l_2^{(\ka)}
       \tanh\left(\frac{\kappa \pi}{4}\right)\,, \\
(M_3)_{2m,2n}=&
\int_0^\infty d\kappa
 \frac{l_3^{(\kappa)}}{\N(\kappa)} \frac{4}{\theta^2}
   \sum_{k,l=1}^\infty T_{2m,2l-1}T_{2n,2k-1}
     \frac{v_{2l-1}^{(\kappa)}v_{2k-1}^{(\kappa)}}{\sqrt{(2l-1)(2k-1)}}= \\
=&
\frac{4}{\theta^2}\int_0^\infty d\kappa
 \frac{v_{2m}^{(\ka)}v_{2n}^{(\ka)}}{\N(\ka)\sqrt{2m}\sqrt{2n}} l_3^{(\ka)}
       \tanh^2\left(\frac{\kappa \pi}{4}\right)\,.
\end{split}
\end{equation}
In the case of $M_2,M_3$ we also used the identity
\begin{equation}
\sum_{k=1}^\infty T_{2n,2k-1}\frac{v_{2k-1}^{(\kappa)}}{\sqrt{2k-1}}=
-\frac{v_{2n}^{(\kappa)}}{\sqrt{2n}}\tanh\left(\frac{\kappa \pi}{4}\right)\,,
\end{equation}
which is an immediate consequence of eq.~(6.7,6.8,3.10) of \cite{Douglas:2002jm}.

In \cite{Bars:2002nu} Bars and Matsuo have considered the subalgebra of (shifted)
Gaussian states. This subalgebra has a form of a monoid, when the
Gaussian matrices are not restricted to be
positive, and the star-multiplication is formally defined in the same way for all
the matrices.
They have found that a state $M$ in the monoid is a projector if either $M=0$,
in which case it represents the identity, or
\begin{equation}
\label{BMEq}
(M\sigma)^2=\One\,,
\end{equation}
where $\sigma$
is the noncommutativity matrix, given by
\begin{equation}
\sigma=i\theta\mat{\phantom{-}0 & \One \\ -\One & 0}\,.
\end{equation}
The condition $(M\sigma)^2=\One$ implies that $\theta M$ is an element of
$SP(\infty)$.
This reminds us of the projector criteria we discussed at section \ref{contMoyal}.
Of course, one has to well define the meaning of $SP(\infty)$, according to the
class of matrices which are considered legitimate.

In terms of the matrices $M_1,M_2,M_3$, the condition~(\ref{BMEq}) tells that the matrices
$M_2M_1$ and $M_3M_2$ are symmetric, and that
\begin{equation}
M_1M_3-M_2^2=\One\,.
\end{equation}
For states in $\HH_{\ka^2}$ the matrices $M_2M_1$ and $M_3M_2$ are always
symmetric, regardless of the projector condition. As for the last condition, it
is an immediate result of the form of $M_1,M_2,M_3$, eq.~(\ref{m1m2m3}),
of the orthogonality condition~(\ref{orthoEv}), and of the projector condition,
eq.~(\ref{moyalCond}).
We see that our results are indeed compatible with the general constraint,
eq.~(\ref{BMEq}).

\section{Conclusions}

In this paper we set out to find squeezed state projectors
whose matrix commute with $(\K_1)^2$.
The set $\HH_{\ka^2}$ of squeezed states that commute with $(\K_1)^2$
is a subalgebra of the star algebra.
Analyzing this subalgebra is straightforward, using the explicit
form of $V_3$ (\ref{V3}).
This subalgebra obviously contains the wedge states,
$\HH_{\it wegde}\subset\HH_{\ka^2}$.
The generalized butterfly states are also in $\HH_{\ka^2}$,
but some other surface states are not.
There are also states in $\HH_{\ka^2}$ which are not surface states.

All this makes $\HH_{\ka^2}$ a convenient laboratory for the study of
the star algebra.
The BPZ reality condition is given by eq.~(\ref{SReal}).
The condition for twist invariance is then
that the matrix is real, eq.~(\ref{TwistInv}).
The projection condition also has a simple form, eq. (\ref{ManySolutions}).

The $\HH_{\ka^2}$ subalgebra has a very simple representation in
the continuous Moyal formalism (\ref{tildeS}).
Using this formalism we found that all the projectors in $\HH_{\ka^2}$
have the same (divergent) normalization (\ref{ProjNorm}).
We used this normalization to show that any two generalized butterfly
states are orthogonal (\ref{ButtOrth}).

It would be interesting to try to address some of the open problems
of string field theory in the $\HH_{\ka^2}$ subalgebra.
\begin{itemize}
\item If all rank one projectors represent the same $D$-brane they should
be related by a gauge transformation. The fact that all the projectors
in $\HH_{\ka^2}$ have the same normalization and that we know what is the
symmetry that relates them to each other, could be a step in the right
direction.

\item The allowed space of states in SFT is not well understood.
In the $\HH_{\ka^2}$ subalgebra this question should translate to
conditions on the functions of $\ka$.
Those functions must be integrable to have any meaning.
But we do not know what other constraints they have and
specifically we do not know if we should impose continuity.

\end{itemize}

Addressing the above questions would require an analysis of the
ghost sector and of the zero-modes.
These two issues were neglected in this paper.
It might be possible to incorporate these issues in
the $\HH_{\ka^2}$ subalgebra using the known spectrum of the full
vertex.

\acknowledgments

We are in much debt to Leonardo Rastelli for illuminating discussions,
and many suggestions.
We would also like to thank Jacob Sonnenschein, Yaron Oz, Akikazu Hashimoto,
Yaron Kinar and Ofer Aharony
for fruitful discussions.
MK would like to thank the institute of advanced study at Princeton
where part of this work was done, for hospitality.
This work was supported in part by the US-Israel Binational Science
Foundation, by the German-Israeli Foundation for Scientific Research,
and by the Israel Science Foundation. 

\appendix
\section{Butterfly spectroscopy}

In this appendix we show, independently of the methods developed in the
main text, that the butterfly state~\cite{Gaiotto:2001ji} shares
eigenvectors with the wedge states and compute its eigenvalues.

In the oscillator representation the butterfly is given as a squeezed state
with the following Neumann coefficients
\begin{equation}
\label{Vmat}
V^B_{mn}=\begin{cases}-(-1)^\frac{m+n}{2}\frac{\sqrt{mn}}{m+n}
    \frac{\Gamma(\frac{m}{2})\Gamma(\frac{n}{2})}
         {\pi \Gamma(\frac{m+1}{2})\Gamma(\frac{n+1}{2})} & m,n \text{ odd}\\
    0 & \text{otherwise}\end{cases}\,.
\end{equation}

In \cite{Rastelli:2001hh} the spectrum of the wedge states $T_n$ was found
based on the commutation relation
\begin{equation}
\label{TcomK}
[T_n,\K_1]=0\,,
\end{equation}
and the non-degeneracy of $\K_1$'s spectrum. Yet,
\begin{equation}
[V^B,\K_1]\neq 0\,,
\end{equation}
Thus our claim that the butterfly shares 
eigenvectors with the family of wedge states requires an explanation.
The way out is to note that although $\K_1$ is non-degenerate $T_n$ are
doubly degenerate.
Therefore, there are two sets of commuting operators that can be 
diagonalized simultaneously.
The first set includes $\K_1$ and $T_n$
while the second includes $(\K_1)^2, T_n$ and $C$. We need to show that $C$ and
$T_n$ commute with $V^B$ in order to prove our claim, since $C$ lifts the 
degeneracy of the eigenvectors of $T_n$.
From the explicit form of $V^B$ as given in (\ref{Vmat}) we see that $V^B$
is twist invariant
\begin{equation}
\label{VcomC}
[V^B,C]=0\,,
\end{equation}
and that all its twist even eigenvalues vanish
\begin{equation}
\label{Veven}
V^B v_+^{(\ka)}=0\,.
\end{equation}
The squeezed states $T_n$ can be represented as functions of a single
matrix $B$ defined by
\begin{equation}
T_{2+\e}=\e B+O(\e^2)\,,
\end{equation}
which was introduced in~\cite{Rastelli:2001hh}. Its entries are
\begin{equation}
\label{Bmat}
B_{mn}=\begin{cases}
    -\frac{(-1)^\frac{n-m}{2}\sqrt{mn}}{(m+n)^2-1} & m+n\text{ even}\\
    0 & \text{otherwise}\end{cases}\,.
\end{equation}
The twist odd eigenvalues of $B$ are non-degenerate.

In section~\ref{secVcomB} we will rely on the simple form of the matrix $B$
to prove that $[V^B,T_n]=0$.
We shall use this result in section~\ref{secBflyEigen} in order to find the 
eigenvalues of the butterfly.

\subsection{The butterfly commutes with the B matrix}
\label{secVcomB}

Both $B$ and $V^B$ are symmetric matrices. Two such matrices commute iff their
product $G \equiv BV^B$ is a symmetric matrix. It is immediate that $G_{mn}=0$
unless both $m,n$ are odd. Now we use eq.
(\ref{Bmat},\ref{Vmat}) to find
\begin{multline}
\label{gEq}
G_{mn}=\sum_{l=1}^\infty B_{ml}V^B_{ln}=\\
  \frac{2 \sqrt{mn} (-1)^\frac{n-m}{2}\Gamma(\frac{n}{2})}
       {\pi \Gamma(\frac{n+1}{2})}
\sum_{k=1}^\infty
    \frac{\Gamma(k+\frac{1}{2})}{\Gamma(k)(2k+n-1)((m+2k-1)^2-1)}\,,
\end{multline}
where in the second line we used the fact that $V^B_{ln}=0$ unless $l=2k-1$,
in order to change the summation from $l$ to $k$.

To evaluate this sum we multiply the summands by $x^k$.
The sum converges for $0<x<1$, and we shall see that the limit
$x \rightarrow 1$ exists. We rewrite the sum as
\begin{multline}
\sum_{k=1}^\infty \frac{\Gamma(k+\frac{1}{2})}
                       {\Gamma(k)(2k+n-1)(2k+m-2)(2k+m)}x^k=\\
 \sum_{k=1}^\infty \frac{\Gamma(k+\frac{1}{2})}{\Gamma(k)}x^k
  (\frac{a}{2k+n-1}+\frac{b}{2k+m-2}+\frac{c}{2k+m})\,,
\end{multline}
where
\begin{equation}
a=\frac{1}{m^2-2mn+n^2-1}\,,\quad
b=\frac{1}{2-2m+2n}\,,\quad
c=\frac{1}{2+2m-2n}\,.
\end{equation}

For $x<1$ we can change the summation order and sum each term separately.
All three sums are of the same form. A general sum of this form gives
\begin{multline}
\sum_{k=1}^\infty\frac{\Gamma(\frac{1}{2} + k)x^k}{\Gamma(k)(k+u)}=\\
-\frac{\sqrt{\pi}((2u(x-1)+x-2)B_x(u+1,\frac{1}{2})+
    (3+2u)B_x(u+1,\frac{3}{2}))}{2(x-1)x^u}\,,
\end{multline}
where $B_x(u+1,\frac{1}{2})$ is the incomplete beta function.
Expanding this term using $x=1-\e$ gives
\begin{equation}
-\sqrt\frac{\pi}{\e}-\frac{\pi\Gamma(u+1)}{\Gamma(u+\frac{1}{2})}+O(\e)\,.
\end{equation}
Now we substitute
\begin{equation}
u=\frac{n-1}{2},\ \frac{m-2}{2},\ \frac{m}{2}\,,
\end{equation}
multiply by $a,b,c$, and sum the three terms.
The singular terms drop out since
\begin{equation}
a+b+c=0,
\end{equation}
and we are left with
\begin{equation}
-\frac{\pi}{2}
\left(
\frac{\Gamma(\frac{n+1}{2})}{((m-n)^2-1)\Gamma(\frac{n}{2})}-
\frac{\Gamma(\frac{m}{2})}{\Gamma(\frac{m+1}{2})}\frac{m+n-1}{4((m-n)^2-1)} 
\right)\,.
\end{equation}
Plugging this result into eq. (\ref{gEq}) we get two terms which
are symmetric with respect to interchanging $m,n$.
This completes the proof that $[V^B,B]=0$.
Since all $T_n$ are given functions of $B$ we also get
\begin{equation}
[V^B,T_n]=0\,.
\end{equation}

\subsection{The eigenvalues of the butterfly}
\label{secBflyEigen}

After showing that the eigenvectors of $V^B$ are the same as those of $B$,
we turn to find the corresponding eigenvalues.
All the twist even eigenvectors have a zero eigenvalue (\ref{Veven}).
For the twist odd eigenvectors it is enough to calculate
\begin{equation}
\label{eigen}
\lambda_{v_-^{(\ka)}}=\frac{1}{v_{-,m}^{(\ka)}}\sum_{n=1}^\infty
    V_{mn}v_{-,n}^{(\ka)}\,,
\end{equation}
for any $m$ such that $v_{-,m}^{(\kappa)}\neq 0$.
We choose to perform the calculation with $m=1$.
$v_-^{(\kappa)}$ is given by the generating function \cite{Rastelli:2001hh}
\begin{equation}
\label{vGF}
\sum_{m=1}^\infty \frac{v_{-,m}^{(\kappa)}}{\sqrt m}z^m =
 \frac{1}{\kappa}{\sinh(\kappa \tan^{-1}(z))}\,,
\end{equation}
with the inverse relation
\begin{equation}
\label{vGFinv}
v_{-,m}^{(\kappa)}=\frac{\sqrt{m}}{2 \pi i \kappa}\oint
 \frac{\sinh(\kappa \tan^{-1}(z))}{z^{m+1}} dz\,,
\end{equation}
where $z$ is a small contour around the origin.

We can now calculate the eigenvalues of eq. (\ref{eigen}) setting $m=1$
\begin{equation}
\label{eigen2}
\begin{split}
\lambda_{v^{(\kappa)}}
&=\frac{1}{v_{-,1}^{(\kappa)}}\sum_{k=1}^\infty (-1)^{k+1}\frac{\sqrt{2k-1}}{2k}
 \frac{\Gamma(\frac{2k-1}{2})}{\sqrt \pi \Gamma(k)}
 v_{-,2k-1}^{(\kappa)}\\
&=\sum_{k=1}^\infty (-1)^{k+1}
 \frac{\Gamma(k+\frac{1}{2})}{\sqrt \pi \Gamma(k+1)}
 \oint \frac{ \sinh(\kappa \tan^{-1}(z))}{2\pi i \kappa z^{2k}} dz\,,
\end{split}
\end{equation}
where in the first equality we set $n=2k-1$, and use the expression for the
matrix $V^B$ (\ref{Vmat}).
In the second equality we use the generating function (\ref{vGFinv}).

We would like to change the order of summation and integration.
For that we need $|z|>1$, but the generating function has two branch cuts,
going from $i$ to $i\infty$, and from $-i$ to $-i \infty$. Therefore we deform
the contour as in figure \ref{fig:contour}. The contour is now given by the
integrals on the small semi-circles with radius $\epsilon$
(marked A in the figure),
the large semi-circles with radius $R$ (B), and the straight
lines around the cuts $(\mbox{C}^\pm)$.
The contribution to the integral of the parts
A and B goes to zero as $\epsilon \rightarrow 0$ and $R \rightarrow \infty$.
Both integrals on $\mbox{C}^+$ contribute the same, and the same goes for
both $\mbox{C}^-$ integrals.
We can now write the integral as
\begin{multline}
\oint \frac{\sinh(\kappa \tan^{-1}(z))}{2\pi i \kappa z^{2k}} dz=
2 \lim_{\epsilon \rightarrow 0}\lim_{R \rightarrow \infty}
 \int_{\mbox{C}^+ + \mbox{C}^-}
 \frac{\sinh(\kappa \tan^{-1}(z))}{2\pi i \kappa z^{2k}} dz=\\
\lim_{\epsilon \rightarrow 0^+}\int_1^\infty
 \frac{\sinh(\kappa \tan^{-1}(i x-\epsilon))-
  \sinh(\kappa \tan^{-1}(i x+\epsilon))}
   {\pi \kappa (i x)^{2k}}dx=\\
\frac{(-1)^{k+1}}{\pi \kappa}\int_1^\infty
 \frac{\sinh(\kappa ( \frac{\pi}{2}+i \coth^{-1}(x)))-
       \sinh(\kappa (-\frac{\pi}{2}+i \coth^{-1}(x)))}
   {x^{2k}}dx=\\
\frac{2(-1)^{k+1} \sinh\left(\frac{\kappa \pi}{2}\right)}{\pi \kappa}\int_1^\infty
 \frac{
       \cos(\kappa \coth^{-1}(x))}
   {x^{2k}}dx\,.
\end{multline}

\FIGURE[hbt]{
\centerline{\begin{picture}(0,0)%
\epsfig{file=Contour.pstex}%
\end{picture}%
\setlength{\unitlength}{1973sp}%
\begingroup\makeatletter\ifx\SetFigFont\undefined%
\gdef\SetFigFont#1#2#3#4#5{%
  \reset@font\fontsize{#1}{#2pt}%
  \fontfamily{#3}\fontseries{#4}\fontshape{#5}%
  \selectfont}%
\fi\endgroup%
\begin{picture}(7244,8444)(1179,-8183)
\put(5401,-1861){\makebox(0,0)[lb]{\smash{\SetFigFont{6}{7.2}{\familydefault}{\mddefault}{\updefault}C$^-$}}}
\put(5401,-6361){\makebox(0,0)[lb]{\smash{\SetFigFont{6}{7.2}{\familydefault}{\mddefault}{\updefault}C$^+$}}}
\put(8101,-1861){\makebox(0,0)[lb]{\smash{\SetFigFont{6}{7.2}{\rmdefault}{\mddefault}{\updefault}B}}}
\put(1501,-1861){\makebox(0,0)[lb]{\smash{\SetFigFont{6}{7.2}{\rmdefault}{\mddefault}{\updefault}B}}}
\put(4501,-4861){\makebox(0,0)[lb]{\smash{\SetFigFont{6}{7.2}{\rmdefault}{\mddefault}{\updefault}A}}}
\put(5101,-3061){\makebox(0,0)[lb]{\smash{\SetFigFont{6}{7.2}{\rmdefault}{\mddefault}{\updefault}A}}}
\put(4876,-5311){\makebox(0,0)[lb]{\smash{\SetFigFont{6}{7.2}{\familydefault}{\mddefault}{\updefault}$\epsilon$}}}
\put(3901,-1861){\makebox(0,0)[lb]{\smash{\SetFigFont{6}{7.2}{\familydefault}{\mddefault}{\updefault}C$^+$}}}
\put(3901,-6361){\makebox(0,0)[lb]{\smash{\SetFigFont{6}{7.2}{\familydefault}{\mddefault}{\updefault}C$^-$}}}
\put(6751,-2461){\makebox(0,0)[lb]{\smash{\SetFigFont{6}{7.2}{\rmdefault}{\mddefault}{\updefault}R}}}
\end{picture}
}
\caption{The original contour of integration (the small black circle) is
replaced by the more complicated path
that satisfies $|z|>1$ except for the segment A.
in a way that avoids the branch cuts.
Only the blue lines ($\mbox{C}^\pm$) contribute in the limit
$R\rightarrow\infty,\epsilon\rightarrow0$.}
\label{fig:contour}
}

Now we can change the order of summation and integration in
eq. (\ref{eigen2}) to get
\begin{equation}
\label{ButtEigen}
\begin{split}
\lambda_{v_-^{(\kappa)}}
&=\frac{2\sinh(\frac{\kappa \pi}{2})}{\pi^{3/2} \kappa}\int_1^\infty
       \cos(\kappa \coth^{-1}(x))\sum_{k=1}^\infty
 \frac{\Gamma(k+\frac{1}{2})}{\Gamma(k+1) x^{2k}} dx\\
&=\frac{2\sinh(\frac{\kappa \pi}{2})}{\pi \kappa}\int_1^\infty
       \cos(\kappa \coth^{-1}(x))
(\frac{x}{\sqrt{x^2-1}}-1) dx\\
&=\frac{2\sinh(\frac{\kappa \pi}{2})}{\pi \kappa}\int_0^\infty
       \frac{\cos(\kappa u)}{1+\cosh(u)} du=
        \frac{1}{\cosh(\frac{\kappa \pi}{2})}\,.
\end{split}
\end{equation}
These are the twist odd eigenvalues of the butterfly.
This result is in agreement with the results of the main text.

\bibliography{FKM}

\end{document}